\documentclass[showpacs,twocolumn]{revtex4}
\usepackage[english]{babel}
\usepackage[utf8]{inputenc}
\usepackage{graphicx}
\usepackage{amsmath, amssymb}
\usepackage{dcolumn}
\usepackage{subfigure}
\usepackage{slashed}
\usepackage{feynmf}

\newcommand{\be}{\begin{eqnarray}}
\newcommand{\ee}{\end{eqnarray}}

\newcommand{\m}{\mathcal}
\newcommand{\ik}{\int_k}
\newcommand{\iq}{\int_q}

\newcommand{\tr}{\text{Tr}\,}

\begin{document}

\title{\bf On the Bose symmetry and the left- and right-chiral anomalies}

\author{J. S. Porto$^{(a)}$}\email[]{joilsonp@ufmg.br}
\author{A. R. Vieira$^{(b)}$}\email[]{alexandre.vieira@uftm.edu.br}
\author{A. L. Cherchiglia$^{(c)}$}\email[]{adriano.cherchiglia@ufabc.edu.br}
\author{Marcos Sampaio$^{(a)(c)}$}\email[]{msampaio@fisica.ufmg.br}
\author{Brigitte Hiller$^{(d)}$}\email[]{brigitte@teor.fis.uc.pt}

\affiliation{(a) Universidade Federal de Minas Gerais - Departamento de F\'{\i}sica - ICEX \\ P.O. BOX 702,
30.161-970, Belo Horizonte MG - Brazil}
\affiliation{(b) Universidade Federal do Tri\^angulo Mineiro - Campus Iturama, 38280-000, Iturama MG - Brazil}
\affiliation{(c) Universidade Federal do ABC - Centro de Ciências Naturais e Humanas \\ 09210-580, Santo Andr\'e SP - Brasil.}
\affiliation{(d) CFisUC, Department of Physiscs, University of Coimbra, 3004-516 Coimbra, Portugal}

\begin{abstract}

It is generally assumed that in order to preserve Bose symmetry in the left- (or right-chiral) current it is necessary to equally 
distribute the chiral anomaly between the vectorial and the axial Ward identities, requiring the use of counterterms to restore consistency. In this work, we show how to calculate the quantum breaking of the left- and 
right-chiral currents in a way that allows to preserve Bose symmetry independently of the chiral anomaly, using the Implicit Regularization method.

\end{abstract}

\maketitle

\section{Introduction}\label{s1} 

Almost half a century after the seminal papers \cite{Jackiw:1969, Adler:1969}, reporting the existence of theories with anomalous breaking of symmetries, it remains to present date still a challenge to deal explicitly with these quantum breakings in a perturbative field theoretical approach,  as the latter requires the use of an invariant regularization and renormalization program. 

The necessity to have the gauge symmetries of the Standard Model implemented to all loop orders led to the advent of dimensional regularization (DR) \cite{DR, Bollini}, which has been  since then one of the most widely used regularizations, as it also respects unitarity and causality to all orders. DR is however burdened by the technical complexity associated in dealing with dimension specific objects,  such as the Levi-Civita tensor and the $\gamma_5$ matrix \cite{Siegel:1979, Siegel:1980, Jegerlehner}, which can be present in theories with anomalous breakings. Several works have been dedicated to devise methods capable of handling these structures to yield unambiguous results in the limit of recovering the physical dimension of a theory \cite{Tsai1:2011, Tsai2:2011, Ferrari:2014, Ferrari:2015, Ferrari:2016}. 

The importance of being able to focus directly on the Ward identities in renormalized perturbation calculations in gauge theories with chiral fermions has been emphasized in \cite{Cheng} and requires a careful treatment of the $\gamma_5$ matrix. 

The correct assessment of Bose symmetry and anomalous processes is of pertinent interest for the study of certain hadronic processes in current experimental facilities, see \cite{Osipov:2017}.

Recently we became aware that even if one stays in the dimension of the theory, certain operations involving the $\gamma_5$  algebra may lead to ambiguous results if used in divergent integrals \cite{Viglioni2016}. In particular the use of the anticommutator $\{\gamma_\mu,\gamma_5\}=0$ has been identified as a source of ambiguities. The root of this problem seems to reside in illegal symmetric integrations in the momentum variable of a divergent integral \cite{perez:2001, inprogress}. We have forwarded a minimal prescription to deal with $\gamma_5$ that leads to unambiguous results; it simply relies in using its definition (equivalent to symmetrization of the trace) \cite{Vieira, Cynolter:2011, Ma-Wu,  Perez}  and avoiding the use of the anticommutator. To show this, use has been made of the Implicit Regularization (IReg) framework \cite{IR,BaetaScarpelli:2000zs,Cherchiglia:2010yd}, which allows to keep open the choice of the Ward Identity (WI) to be (or not) satisfied, until the very end of the evaluation of an amplitude allowing for a democratic display of physical anomalies. This is achieved in terms of a set of arbitrary surface terms that are extracted as differences of basic divergent integrals (BDI) (independent on physical properties, such as masses and external momenta) of the same degree of divergence. This approach reflects the matters raised by Jackiw in \cite{Jackiw:2000} about the occurrence of finite but arbitrary parameters which appear in certain perturbative calculations and are regularization dependent. In addition, it should be emphasized that the choice of the WI to be satisfied (or not) can be made without breaking momentum routing invariance of an amplitude \cite{Viglioni2016, Vieira}.

In this contribution we use IReg as a tool to identify and resolve the source of conflict in the implementation of Bose symmetry in the left-right (chiral) sector of gauge currents, that reportedly \cite{Pokorski} constrains the values of the anomaly in the related vector - axial representation.  For that purpose we use  effective vertices of the abelian chiral Schwinger model in two dimensions (2-d)  and the Adler-Bell-Jackiw (ABJ) triangle anomaly \cite{Jackiw:1969, Adler:1969} and show  that the conflict in these cases can also be traced back to the incautious use of the property   
$\{\gamma_\mu,\gamma_5\}=0$ in divergent amplitudes when we choose to relate axial and vectorial vertices with right and left fields. In fact the anti-commutator evaluates, in the cases considered in the present contribution, all chirality mixing amplitudes to zero, prior to the use of a regularization. By avoiding this relation and using IReg combined with the symmetrization of the trace, one obtains full consistency of the Bose symmetry and the values that the anomaly can take in the equivalent left-right and vector-axial representations. 

The paper is organized as follows. After a resum\'ee of the IReg in the next chapter, we present in section \ref{s2} the quantities to be evaluated in 2-d and 4-d, then proceed to calculate them in sections   \ref{s3-sub1} and \ref{s3-sub2}, and study their anomalous breakings in the light of Bose symmetry. Finally we present our conclusions in \ref{sec4}.

\section{Calculational framework: the Implicit Regularization method}

In this section we present, briefly, the regularization method used to deal with the divergences that will appear in the course of the calculations. Since our aim is to discuss ambiguities related to dimension specific objects (such as the $\gamma_{5} $ matrix), regularization methods based on dimensional continuation are \textit {a priori} inappropriate as the Clifford algebra is ambiguous under dimensional continuation. Therefore, regularization methods that operate in the physical dimensions of the underlying quantum field theory seem more appealing to tackle the problem at hand. For a recent review of different methods, see \cite{Gnendiger:2017pys}. 

In this work, we adopt the framework of IReg whose basic characteristic is to use an algebraic identity in divergent integrands in order to isolate divergences as unevaluated integrals free of physical quantities (physical momenta or masses, in general). The divergent integrals thus generated can be further simplified to a well-defined set of scalar integrals only, at the expense of regularization-dependent objects (surface terms). It had been shown that the latter are at the root of violation of abelian gauge symmetry \cite{Adriano}, and must be set to zero in this case, while it is conjectured that they are related to symmetry breakings in general.

For ease of the reader, we illustrate the method reviewing some of the results of \cite{Viglioni2016} which will be used in the next sections. In this reference the chiral Schwinger model was studied and the one-loop two-point function of the photon with a vectorial and a chiral vertex was computed. It's amplitude is given by
\be
\Pi_{\mu\nu}^{VA}=-iTr\int_{q}\gamma_{\mu}\frac{1}{\slashed{q}-\slashed{p}}\gamma_{\nu}\gamma_{5}\frac{1}{\slashed{q}},
\label{eq.PiVA}
\ee
where $\int_{q}$ stands for $\int \frac{d^{2}q}{(2\pi)^{2}}$. To evaluate this integral according to IReg one must just follow the steps:
\vspace{0.5cm}

\begin{enumerate}
\item Perform Dirac Algebra. As reviewed recently in \cite{Gnendiger:2017pys}, this step is crucial to obtain terms of $q^2$ in the numerator, which should be canceled against propagators. Proceeding otherwise would generate spurious finite terms, invalidating the connection between surface terms and symmetry breakings;
\item Introduce a fictitious mass ($\mu^2$) in propagators to control spurious infrared divergent terms that will appear in the course of the evaluation;
\item Use Eq.~(\ref{eq.Identity}) as many times as necessary to free the divergent integrals from physical quantities (the physical momentum $p$ in the example),
 \begin{equation}
\frac{1}{(q+p)^2-\mu^2} = \frac{1}{q^2-\mu^2} -\frac{(p^2+2p\cdot k)}{(q^2-\mu^2)[(q+p)^2-\mu^2]}.
\label{eq.Identity}
\end{equation}
\item Divergences are now written in terms of a well-defined set of logarithmic BDI's as 
\begin{equation}
I^{\mu_1 \cdots \mu_{2n}}_{log}(\mu^2)\equiv \int_q \frac{q^{\mu_1}\cdots q^{\mu_{2n}}}{(q^2-\mu^2)^{1+n}}.
\end{equation}
In this work only the logarithmic divergences occur, see \cite{Cherchiglia:2014} for a compreensive discussion of the quadratic divergences.
\item Reduce the tensorial BDI's to the scalar one, at the expense of surface terms. For instance,
\begin{eqnarray}
&&g^{\mu\nu}\upsilon_0= \nonumber \\
&&g^{\mu \nu}\int \frac{d^2 q}{(2\pi)^2} \frac{1}{(q^2-\mu^2)}-2\int \frac{d^2 
q}{(2\pi)^2}\frac{q^{\mu}q^{\nu}}{(q^2-\mu^2)^{2}}
\label{eq.TS}
\end{eqnarray}
\item Perform the limit $\mu^2 \rightarrow 0$. In the examples used in this work, the latter will always be well-defined. If this was not the case, a scale $\lambda^2 \neq 0$ would be introduced which would play the role of the renormalization scale in the renormalization group equations.  
\end{enumerate}
\vspace{0.5cm}
This program can be successfully extended to higher loop orders \cite{Cherchiglia:2010yd, Pontes, Dias}.

Therefore, applying the steps above to Eq.~(\ref{eq.PiVA}) gives
\be
\Pi_{\mu\nu}^{VA}=-2i\epsilon_{\nu\theta}\left[\frac{(\delta_{\mu}^{\theta}p^{2}-p_{\mu}p^{\theta})(-2b)}{p^{2}}-\delta_{\mu}^{\theta}\upsilon_{0}\right]=\Pi_{\nu\mu}^{AV}
\label{first}
\ee
where $b=i/4\pi$, and $\upsilon_{0}$ is the surface term defined in Eq.~(\ref{eq.TS}) which is ambiguous. From the result above, the two Ward identities, the vectorial and axial one, are easily obtained:
\begin{align}
p^{\mu}\Pi_{\mu\nu}^{VA}=2i\epsilon_{\nu\theta}p^{\theta}\upsilon_{0},\nonumber\\
p^{\nu}\Pi_{\mu\nu}^{VA}=-2i\epsilon_{\mu\theta}p^{\theta}(2b+\upsilon_{0}).
\label{WIS}
\end{align}

We notice the appearance of an arbitrariness parametrized as a surface term which allows a democratic view of the anomaly. Similar analysis can be performed for the other components of the complete two-point function in the Chiral Schwinger Model, namely $\Pi_{\mu\nu}^{VV}$ and $\Pi_{\mu\nu}^{AA}$, with the result
\be
\Pi_{\mu\nu}^{VV}=\Pi_{\mu\nu}^{AA}=-2i\left[\frac{(g_{\mu\nu}p^{2}-p_{\mu}p_{\nu})(-2b)}{p^{2}}-g_{\mu\nu}\upsilon_{0}\right].
\label{AAVV}
\ee

Finally, we emphasize that the procedure described in this section is not restricted to 2-d theories, rather it can be easily generalized to arbitrary (integer) dimensions. For instance, the surface term $\upsilon_{0}$ will be thus defined  in 4-d as
\begin{equation}
g^{\mu\nu}\upsilon_0=g^{\mu \nu}\int \frac{d^4 q}{(2\pi)^4} \frac{1}{(q^2-\mu^2)^2}-4\int \frac{d^4 
q}{(2\pi)^4}\frac{q^{\mu}q^{\nu}}{(q^2-\mu^2)^{3}},
\end{equation} 
and similar considerations also hold for the BDI's.

IReg is compatible with the recursion formula of the Bogoliubov, Parasiuk, Hepp, Zimmermann (BPHZ) framework, \cite{BPHZ}, as discussed in \cite{Sampaio:2002ii, Cherchiglia:2010yd}, therefore it respects unitarity, causality and Lorentz invariance, basic requisites that any regularization method should fullfill.  IReg was developped having additionally in mind that the symmetries of the underlying theory be fulfilled at the largest extent possible in the process of regularization. In this respect  IReg proves to be superior to the original BPHZ scheme. For instance, it is well known that the BPHZ scheme, although providing a recipe to obtain finite quantities, breaks non-abelian gauge invariance. IReg provides for a means to reconstruct the symmetries through the  above-mentioned systematic classification of the surface terms. We reiterate that these are finite and regularization dependent quantities which are {\it a priori} arbitrary valued.   If one decides, for instance, to preserve vectorial gauge symmetry, it is just necessary to set all of the surface terms to zero. However the method is sufficiently general  that no choice regarding symmetries must be performed until the very end of the computation, as all regularization dependent information can be kept arbitrary until then.  The symmetry content of the theory dictates at the end the values that the arbirary parameters must take. This fact offers a neat arena to discuss anomalies, with the application of the method in different contexts
\cite{BaetaScarpelli:2001ix,Souza:2005vf,Ottoni:2006ij,Scarpelli:2008fw,Felipe:2011rs,Adriano,Cherchiglia:2012zp,Felipe:2014gma,Vieira,Viglioni2016}.

A different framework is handled in the calculation of anomalous processes,  which uses the BPHZ renormalized form for the master equation of the field antifield quantization \cite{BV} proposed by De Jonghe,
Paris and Troost \cite{Paris}.  Initially developed to address gauge anomalies, this framework can be adapted  to deal with global anomalies too, and generates the Ward Identities associated with the axial anomaly \cite{Abreu}.    
\\

\section{Overview on the quantum breaking of classical currents}\label{s2}

A $d-$dimensional action for massless fermions interacting via general axial and vectorial couplings can be written as:
\be
S=\int d^dx\ \bar{\psi}(i\slashed{\partial}+e\slashed{V}+e\slashed{A}\gamma_5)\psi,
\label{eq1}
\ee
where $V^{\mu}$ and $A^{\mu}$ stand for a general external vector and axial field, respectively.  QED is recovered when $A^{\mu}=0$ and  chiral
QED when $A^{\mu}=V^{\mu}$.

\noindent The classical action of $(\ref{eq1})$ is invariant under both the local gauge and local axial transformations  $U_V(1)\times U_A(1)$,  with associated conserved classical   
currents
\begin{eqnarray}
&\partial_{\mu}j^{\mu}_V=\partial_{\mu}(\bar{\psi}\gamma^\mu\psi)=0, \nonumber\\
&\partial_{\mu}j^{\mu}_A=\partial_{\mu}(\bar{\psi}\gamma^5\gamma^\mu\psi)=0 .
\label{eq2}
\end{eqnarray}

\noindent By decomposing the fermion field in  Eq. $(\ref{eq1})$ as a sum of  left- and  right-chiral fields,  $\psi=\psi_L+\psi_R$, where 
$\psi_{R,L}=\frac{1}{2}(1\pm \gamma_5)\psi$ are the fields with positive $(R)$ and negative $(L)$ chirality,  the lagrangian of $(\ref{eq1})$ is seen to be also invariant
under left- and right-chiral local gauge transformations,  it is $U_L(1)\times U_R(1)$ symmetric. The left- and right-chiral global gauge transformations 
lead to the left- and right-chiral classical currents
\be
\partial_{\mu}j^{\mu}_{L,R}=\partial_{\mu}(\bar{\psi}_{L,R}\gamma^\mu\psi_{L,R})=0.
\label{eq3}
\ee

In order to determine the anomalous breaking of the symmetry currents $(\ref{eq2})$ and $(\ref{eq3})$ by quantum corrections, one must evaluate the vacuum expectation value of these currents in the
interaction-picture, {\it i.e.} $\langle\partial_{\mu}j^{\mu}_I\rangle$, where $I=V,A,R,L$. For instance, for the axial current in $(1+1)-$dimensions one gets that
\be
\langle\partial_{\mu}j^{\mu}_A\rangle&=&ie\int\ d^2y \partial_{\mu}\Big(i\Pi_{AA}^{\mu\nu}(y)A_{\nu}(x-y)\nonumber\\
&&+i\Pi_{AV}^{\mu\nu}(y)V_{\nu}(x-y)\Big),
\label{eq4}
\ee
where $\Pi_{AV}^{\mu\nu}(y)$ is the two-point Green's function in configuration space,
\be
\Pi_{AV}^{\mu\nu}(y)=\langle0|Tj^{\mu}_A(y)j^{\nu}_V(0)|0\rangle.
\label{eq5}
\ee
The same works for $\Pi_{AA}^{\mu\nu}(y)$ but with two axial currents instead. Therefore, in order to find out the quantum breaking of the
axial current, we need to compute the two-point diagrams $AA$ and $AV$. 
Analogously  the diagrams $VA$ and $VV$ are needed to get information on the quantum
breaking of the vector current, and the diagrams $RR$, $RL$, $LL$
and $LR$ for the anomalous left- and right-chiral
currents.  

The four-divergence of the axial current in $(3+1)-$dimensions is given by	
\begin{widetext}
\begin{align}
&&\langle \partial_\mu j_A^\mu \rangle = -\frac{e^2}{2} \int d^4y d^4z \partial^\mu_{(x)} \Big(T^{AVV}_{\mu\nu\rho}(x,y,z)  V_\nu(x-y) V_\rho(x-z) +T^{AAV}_{\mu\nu\rho}(x,y,z) A_\nu(x-y) V_\rho(x-z)+\nonumber\\
&&+T^{AAA}_{\mu\nu\rho}(x,y,z) A _\nu(x-y) A _\rho(x-z)+T^{AVA}_{\mu\nu\rho}(x,y,z) V _\nu(x-y) A _\rho(x-z) \Big)\,, \nonumber\\
\label{eq:op3d}\end{align}
\end{widetext}

\noindent where $T^{AVV}_{\mu\nu\alpha}$ stands for a three point function with two vector and one axial vertices, for example. 
The anomalous axial current is thus found by analyzing  the diagrams AVV, AAV, AAA and AVA. Similarly the breaking of the vector current requires the evaluation of the VAA, VVA, VVV and VAV  triangle amplitudes.  
\section{Left- and right- chiral versus Axial and vectorial anomalies}
\label{s3}

\subsection{Two dimensional case}\label{s3-sub1}

Consider the two-point function
\begin{eqnarray}
\Pi^{I J}_{\mu\nu} (p) = - i \iq \tr \, \m{V}^{I}_\mu \frac{1}{\slashed{q}-\slashed{p}}\m{V}^{J}_\nu\frac{1}{\slashed{q}}
\end{eqnarray}

\noindent with possible vertices ($ I,J = A,V,L,R $), 
\begin{eqnarray}
\m{V}^V_\mu =\gamma_\mu \,,\quad \m{V}^A_\mu&=&\gamma_\mu \gamma_5 \,,\quad \m{V}^R_\mu =\frac{1}{2}\gamma_\mu(1+\gamma_5) \nonumber\\
\text{and}\,\quad\m{V}^L_\mu&=&\frac{1}{2}\gamma_\mu(1-\gamma_5) \,.
\label{eq:vertices}
\end{eqnarray}

\noindent In order to simplify the following relations between the amplitudes  
\begin{eqnarray}
\Pi^{RR}_{\mu\nu}&=&\frac{1}{4}\left(\Pi^{VV}_{\mu\nu}+\Pi^{AV}_{\mu\nu}+\Pi^{VA}_{\mu\nu}+\Pi^{AA}_{\mu\nu} \right) \,, \nonumber\\
\Pi^{LL}_{\mu\nu}&=&\frac{1}{4}\left(\Pi^{VV}_{\mu\nu}-\Pi^{AV}_{\mu\nu}-\Pi^{VA}_{\mu\nu}+\Pi^{AA}_{\mu\nu} \right)\,, \nonumber\\
\Pi^{RL}_{\mu\nu}&=&\frac{1}{4}\left(\Pi^{VV}_{\mu\nu}+\Pi^{AV}_{\mu\nu}-\Pi^{VA}_{\mu\nu}-\Pi^{AA}_{\mu\nu} \right) \,, \nonumber\\
\Pi^{LR}_{\mu\nu}&=&\frac{1}{4}\left(\Pi^{VV}_{\mu\nu}-\Pi^{AV}_{\mu\nu}+\Pi^{VA}_{\mu\nu}-\Pi^{AA}_{\mu\nu} \right) \,, \nonumber\\
\label{eq6}
\end{eqnarray} 
we shall resort to two approaches. In the first case we use the identity 
\be
\{\gamma_\alpha,\gamma_5\} = 0 \,,
\label{eq7}\ee
and the property $(\gamma_5)^2=1$, leading to

\begin{equation}
\label{eq8}
\Pi^{AV}_{\mu\nu} = \Pi^{VA}_{\mu\nu}
\end{equation} 
\begin{equation}
\Pi^{AA}_{\mu\nu}=\Pi^{VV}_{\mu\nu}\,,
\label{eq9}\end{equation}

\noindent and thus to the following identities
\begin{eqnarray}
\Pi^{RR}_{\mu\nu}  &=& \frac{1}{2} \Pi^{VV}_{\mu\nu}  + \frac{1}{2}\Pi^{AV}_{\mu\nu} \,, \nonumber\\ 
\Pi^{LL}_{\mu\nu}  &=& \frac{1}{2} \Pi^{VV}_{\mu\nu} - \frac{1}{2}\Pi^{AV}_{\mu\nu} \,, \nonumber\\
\Pi^{RL}_{\mu\nu}  &=& \Pi^{LR}_{\mu\nu}  = 0  \,.
\label{eq10}
\end{eqnarray}

In the second approach we avoid using the identity   (\ref{eq7}) and take instead the 2-d definition of $\gamma_5$ 
\be
\gamma_5= \frac{1}{2!} \epsilon^{\alpha\beta} \gamma_\alpha \gamma_\beta.
\ee
 In this case Eq. (\ref{eq9}) remains unchanged, because it can be obtained independently of such identity, as shown in  \cite{Viglioni2016}, with the result given by Eq. (\ref{AAVV}). On the other hand instead of Eq. (\ref{eq8}) one gets now Eq. (\ref{first})
\begin{eqnarray}
\Pi^{VA}_{\mu\nu}=\Pi^{AV}_{\nu\mu}\,.
\label{eq11}\end{eqnarray}
\noindent As a consequence of  Eq.(\ref{eq11}), one obtains from (\ref{eq6}) the following relations
\be
\Pi^{RR}_{\mu\nu} &=& \frac{1}{2}\Pi^{VV}_{\mu\nu} + \frac{1}{4}\left(\Pi^{VA}_{\mu\nu}+\Pi^{VA}_{\nu\mu} \right)\,,\nonumber\\
\Pi^{LL}_{\mu\nu} &=& \frac{1}{2}\Pi^{VV}_{\mu\nu} - \frac{1}{4}\left(\Pi^{VA}_{\mu\nu}+\Pi^{VA}_{\nu\mu} \right)\,,\nonumber\\
\Pi^{RL}_{\mu\nu} &=& -\Pi^{LR}_{\mu\nu} = \frac{1}{4}\left(\Pi^{VA}_{\nu\mu}-\Pi^{VA}_{\mu\nu}\right)\,.
\label{eq12} \ee 
Two different results have been generated relating the amplitudes in the left-right and axial-vector representations, Eqs.  (\ref{eq10}) and (\ref{eq12}), the first resulting from the use of the anti-commutator, the second not.
These differences have drastic implications on the consistency of Bose symmetry or lack thereof of the results. Indeed, 
one can easily verify that the first approach is not consistent with the Bose symmetry of the RR and LL diagrams. For example, Bose symmetry requires that for diagramm RR  
\be
 \Pi^{RR}_{\mu\nu} (p) =    \Pi^{RR}_{\nu\mu} (-p) \,,    
\label{eq13}\ee
which  means that the amplitude   $\Pi^{VA}_{\mu\nu}$ must be symmetric under the exchange of the Lorentz indices $ \mu \leftrightarrow \nu$ and the momenta  $p \leftrightarrow -p $, according to Eq. (\ref{eq10}). However this fixes the values of the axial and vector WI. This can be seen, combining these constraints  with the relations (\ref{WIS}) for the AV anomaly, yielding 
\be
p^\mu \Pi^{VA}_{\mu\nu}(p) = p^\mu \Pi^{VA}_{\nu\mu}(-p) = p^\mu \Pi^{VA}_{\nu\mu}(p) = -2i b\epsilon_{\nu\theta}p^\theta
\label{eq16}\nonumber\\\ee
which fixes the value of the surface term to be $v_0=-b$.  

As opposed to this,  Bose symmetry of the $RR$ and $LL$ diagrams is fulfilled in the second approach given by Eq. (\ref{eq12}), independently of the value of the anomaly in the axial and vector WI  of the amplitude  $\Pi^{VA}_{\nu\mu}$, as can be easily verified. We also note that the Bose symmetric amplitudes RR and LL in Eq.  (\ref{eq12}) can be reconstructed from the  unsymmetric result (\ref{eq10}),  by taking the average $\frac{1}{2}(\Pi^{RR}_{\mu\nu} + \Pi^{RR}_{\nu\mu})$, provided  the  amplitude $\Pi^{VA}_{\mu\nu}$  on the right hand side is calculated without using the anticommutator.  Then condition (\ref{eq11}) leads to the desired result. In the 4-d case this provides for an essential simplification to obtain  the symmetrized trace.

\noindent The relations  (\ref{eq12}) can be used to calculate explicitely the  left-right WI, starting with  Eqs.  (\ref{first}), (\ref{AAVV}) to obtain the following sum and differences involving the VA amplitudes
\be
\label{sd}
\Pi^{VA}_{\mu\nu}+\Pi^{VA}_{\nu\mu} &=& -4i b\left( p_\nu\epsilon_{\mu\theta}+p_\mu\epsilon_{\nu\theta}\right)\frac{p^\theta}{p^2}\,,\nonumber\\
\Pi^{VA}_{\mu\nu}-\Pi^{VA}_{\nu\mu} &=& -4i b\left(  p_\mu\epsilon_{\nu\theta}-p_\nu\epsilon_{\mu\theta}\right)\frac{p^\theta}{p^2}\nonumber\\
&&-4i \epsilon_{\mu\nu}\left( v_0 +2b \right)\,, 
\ee
\noindent which lead to 
\be
p^\mu \Pi^{RR}_{\mu\nu} &=& i \left( \upsilon_0 p_\nu-b\epsilon_{\nu\theta}p^\theta  \right)\,\nonumber\\
p^\mu \Pi^{LL}_{\mu\nu} &=& i \left( \upsilon_0 p_\nu +b\epsilon_{\nu\theta}p^\theta  \right)\,\nonumber\\
p^\mu \Pi^{RL}_{\mu\nu} &=& -p^\mu \Pi^{LR}_{\mu\nu}= -i \left( b+\upsilon_0 \right)\epsilon_{\nu\theta}p^\theta  \,.
\label{eq17} \ee
Before proceeding let us comment on these WI. Regarding the WI of the Bose symmetric amplitudes RR and LL, one sees  that although the AV amplitude is sensitive to the procedure used in the evaluation of the trace, the average  $\Pi^{VA}_{\mu\nu}+\Pi^{VA}_{\nu\mu}$ is free from any ambiguity, as the arbitrary surface term $v_0$ drops out. Nevertheless the WI of the RR and LL amplitudes remain ambiguous, due to the presence of the VV amplitude in Eq. (\ref{eq12}), which depends on $v_0$. The amplitude mixing chirality $\Pi^{RL}_{\mu\nu}$  does not vanish in general. (This is corroborated by the works of \cite{Banerjee:1995np},\cite{Abreu:1997xg}, where in the chiral decomposition of  the 2-d fermionic determinant leading to a gauge invariant result  the amplitude mixing chirality does not vanish in general. In 4-d \cite{Signer} it has been reported recently that within the framework of dimensional regularization gauge invariance and Bose symmetry are maintained simultaneously in the BM \cite{BM} and FDF \cite{Fazio} approaches, and comply with our statement that chirality mixing amplitudes do not vanish in general.) And its WI is  only null   if $v_0=-b$, i.e. for the case of implementing Bose symmetry on the amplitudes calculated with the anti-commutator of $\gamma_5$. For the sake of completeness we mention that for vanishing surface term $v_0$, the WI related with the VV and AA amplitudes yield zero and the WI for the RR and LL differ by a sign, with the same magnitude as for the  RL amplitude.
\vspace{0.5cm}

With these results and Eq. (\ref{WIS}) one obtains
\begin{eqnarray}
\langle \partial_\mu J_V^\mu \rangle &=& i e\upsilon_0 \epsilon^{\nu\rho}F^{A}_{\nu \rho} - 2 i e \upsilon_0 \partial_\mu V^\mu  \,,\nonumber\\
\langle \partial_\mu J_A ^\mu \rangle &=&   - i e (2b+\upsilon_0)\epsilon^{\nu\rho}F^{V}_{ \nu\rho} - 2 i e \upsilon_0 \partial_\mu A^\mu\,,\nonumber\\
\langle \partial_\mu J_R^\mu \rangle &=& -\frac{ i b e}{2}\epsilon^{\nu\rho}F^{R}_{\nu\rho} - i e\upsilon_0\partial^\nu  R_\nu -i e\frac{b+\upsilon_0}{2}\epsilon^{\mu\nu}F^L_{\mu\nu}\,,\nonumber\\
\langle \partial_\mu J_L^\mu \rangle &=& \frac{i b e}{2}\epsilon^{\nu\rho}F^{L}_{\nu\rho} -i e\upsilon_0\partial^\nu  L_\nu +i e\frac{b+\upsilon_0}{2}\epsilon^{\mu\nu}F^R_{\mu\nu}\,,\nonumber\\
\label{eq18}\end{eqnarray}
\noindent where 

\begin{eqnarray}
 F^{B}_{\mu \nu} \equiv \partial_\mu B_\nu - \partial_\nu B_\mu \,,  \nonumber\\ 
\label{eq19}\end{eqnarray}
\noindent with $B$ standing for any of the external fields considered. They are related
in each representation by
\begin{eqnarray}
R_\mu = V_\mu + A_\mu \,,\quad L_\mu = V_\mu-A_\mu .
\label{eq20}\end{eqnarray}
Comparing the results displayed in Eq. (\ref{eq18}), one sees that they comply with the relations among the currents
\be
J_V^\mu = J^\mu_R+J^\mu_L \,,\quad J_A^\mu= J^\mu_R-J^\mu_L
\label{eq:currentrelations}\ee
showing once again that Eqs. (\ref{eq12}), obtained from the second approach, are correct, and do not fix {\it a priori} the value of the anomaly. 

As is well known,  the chiral Schwinger model  describes an unitary theory with a radiatively induced massive gauge
boson with mass \cite{Jackiw:1985}
\begin{equation}
m^2=\frac{e^2}{\pi} \frac{(\lambda +1)^2}{\lambda}
\end{equation}
with a positive value  $\lambda=- 4 i \pi v_0 > 0$  \cite{Jackiw:1985, Viglioni2016, BaetaScarpelli:2000zs}. If one would take the condition $v_0=-b$, resulting from the first approach, one would get  $\lambda=-1$. This shows a further loophole regarding the blind use of the anti-commutator.

\subsection{Four dimensional case}\label{s3-sub2}

\begin{center}
\begin{figure*}
\includegraphics{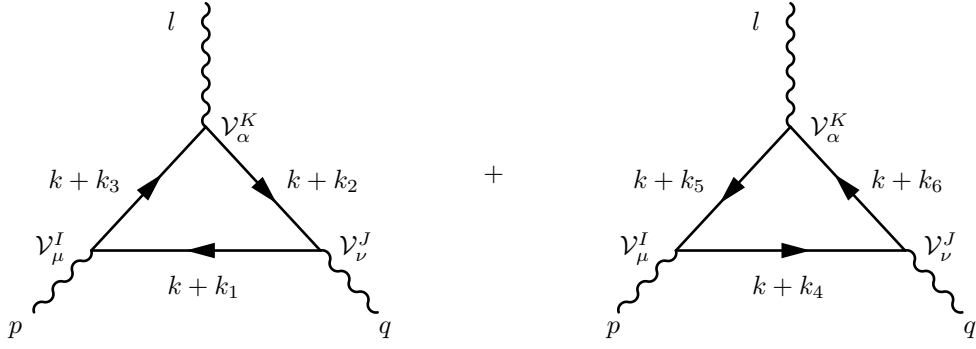}
\caption{The $T^{IJK}_{\mu\nu\alpha}(p,q,l)$ triangle diagram.}
\label{fig:001}\end{figure*}\end{center}

As we have seen in the previous section, the  {\it naive} use of the anti-commuting property  (\ref{eq7}) of the $\gamma_5$ matrix in divergent integrals, leads to inconsistencies in the values of the anomalous quantum breakings of the chiral theory and its equivalent vector-axial formulation, if Bose symmetry is to be satisfied. We thus expect that the inadvertent use of the anti-commutator is also at the root of these inconsistencies in 4-d, which have been reported in the literature \cite{Pokorski}. In this section we show that this is indeed the case.
  
Contributing to the anomaly in 4-d are the 3-point functions

\be
T^{IJK}_{\mu\nu\alpha}(p,q,l) = \int d^4x d^4y d^4z \exp \left(ip x+q y+l z\right)\nonumber\\
\times \langle 0| T j^I_\mu(x) j^J_\nu(y) j^K_{\lambda}(z)|0\rangle \,,
\label{eq21}\ee 
 \noindent where the indices $I,J,K = A,V,L,R $ stand for any combination of the currents, vectorial and axial or left and right, and the external momenta obey 
\be
p+q+l=0 \,.
\label{eq22}\ee 

The 3-point functions, fig.\ref{fig:001}, read in momentum space  
\begin{align}
T^{IJK}_{\mu\nu\alpha}(p,q,l) =-i\ik \text{Tr}\, \m{V}^{I}_\mu \frac{i}{\slashed{k}+\slashed{k_1}} \m{V}^{J}_\nu \frac{i}{\slashed{k}+\slashed{k_2}}\m{V}^{K}_\alpha \frac{i}{\slashed{k}+\slashed{k_3}} \nonumber\\
-i\ik \text{Tr}\, \m{V}^{J}_\nu \frac{i}{\slashed{k}+\slashed{k_4}} \m{V}^{I}_\mu \frac{i}{\slashed{k}+\slashed{k_5}}\m{V}^{K}_\alpha \frac{i}{\slashed{k}+\slashed{k_6}} \,,
\label{eq23}
\end{align}
where 
\be k_2-k_1=q\,,\quad k_3-k_2=l\,,\quad k_1-k_3=p \,,\nonumber\\ k_4-k_6=q\,,\quad k_6-k_5=l\,,\quad k_5-k_4=p\,. \label{eq24}\ee 

\noindent Using Eq.  (\ref{eq7}) and  $(\gamma_5)^2=1$, one obtains the following relations
\be
T^{RRR}_{\mu\nu\alpha}(p,q,l) = \frac{1}{2} T^{VVA}_{\mu\nu\alpha}(p,q,l) \,,\nonumber\\
T^{LLL}_{\mu\nu\alpha}(p,q,l) = -\frac{1}{2} T^{VVA}_{\mu\nu\alpha}(p,q,l)\,,\nonumber\\
T^{AAA}_{\mu\nu\alpha}(p,q,l) = T^{VVA}_{\mu\nu\alpha}(p,q,l)\,,
\label{eq27}\label{eq26}\ee
\noindent where the diagram $T^{VVV}_{\mu\nu\alpha}$ is absent, since it vanishes by the Furry theorem.

Now we proceed to show that these three equations are not consistent with the Bose symmetry pertaining to the diagrams RRR, LLL and AAA. For that we use the WI related with the AVV amplitude  (in the massless case) \cite{Viglioni2016} ,
\begin{subequations}
\begin{gather}
\label{eq:AVVanomaly1}
l^\alpha T^{VVA}_{\mu\nu\alpha}(p,q,l)=-
\frac{1}{2\pi^2}a\epsilon_{\mu\nu\beta\lambda}p^\beta q^\lambda\,,\\
\label{eq:AVVanomaly2}
p^\mu T^{VVA}_{\mu\nu\alpha}(p,q,l)= 
\frac{1}{4\pi^2} (1+a)\epsilon_{\nu\alpha\beta\lambda}p^\beta q^\lambda\,,\\
\label{eq:AVVanomaly3}
q^\nu T^{VVA}_{\mu\nu\alpha}(p,q,l)= 
\frac{1}{4\pi^2} (1+a)\epsilon_{\alpha\mu\beta\lambda}p^\beta q^\lambda\,,
\end{gather}
\label{eq:AVVanomaly}\end{subequations}
\noindent where the parameter $a$ is defined through $\tfrac{1}{4\pi^2}(1+a) \equiv 4i\upsilon_0(\alpha-\beta-1)$, $\upsilon_0$ is again an arbitrary surface term  and $\alpha$, $\beta$ are parameters that are used to specify the  internal momentum routing obeying 
\begin{eqnarray}
k_1=\alpha p +(\beta-1)q\,,\nonumber\\
k_2=\alpha p + \beta q \,,\nonumber\\
k_3=(\alpha-1)p+(\beta-1)q\,.
\end{eqnarray}
\noindent At this point one sees that one has the freedom to chose $a$, either to fulfill  the vector $(a=-1)$ or the axial $(a=0)$ WI. 

Let us now analyse the implications of requiring Bose symmetry. In diagram RRR for example
\be
T^{RRR}_{\mu\nu\alpha} (p,q,l) = T^{RRR}_{\mu\alpha\nu} (p,l,q) = T^{RRR}_{\alpha\nu\mu}(l,q,p)=\dots \,,
\label{eq28}\ee 
and thus  Eq. (\ref{eq26}) imposes by the same symmetry that the $T^{VVA}_{\mu\nu\alpha}$ diagram must be symmetric under subsequent exchange of the same two Lorentz indices and momenta, for instance

\be
\label{exc}
T^{VVA}_{\mu\nu\alpha} (p,q,l) = T^{VVA}_{\mu\alpha\nu} (p,l,q)\, 
\ee 
and thus
\be
\label{ewi}
l^\alpha T^{VVA}_{\mu\nu\alpha} (p,q,l) = l^\alpha T^{VVA}_{\mu\alpha\nu} (p,l,q)  \,.
\ee
\noindent  Comparing the left hand side of this equation with Eq. (\ref{eq:AVVanomaly1}) one sees that it represents the axial WI, and the right hand side is identical to the vector WI, Eq. (\ref{eq:AVVanomaly3}); the latter becomes evident by noting that  Eq. (\ref{exc}) means that the exchange of Lorentz indices and momenta occurs for distinct vertices, i.e. for $V$ and $A$, thus  the vectorial WI is obtained from the right hand side of Eq. (\ref{exc}) by contraction with $l^\alpha$. 
Then, by the momentum conservation  $l=-(p+q)$, 
it reduces to

\be
l^\alpha T^{VVA}_{\mu\alpha\nu}(p,l,q)&=&\frac{1}{4\pi^2}(1+a)\epsilon_{\nu\mu\beta\lambda}p^\beta l^\lambda \nonumber\\
&=& \frac{1}{4\pi^2}(1+a)\epsilon_{\mu\nu\beta\lambda}p^\beta q^\lambda  
 \,.\ee
Consequentially Eq. (\ref{exc}) results in
\be
a=-\frac{1}{3}\,,
\label{eq29}\label{eq:4d-anomaliafixa}\ee
\noindent distributing the anomaly  with equal value in the vector and axial WI.
 
In the following we avoid the identity  (\ref{eq7}) in divergent integrals, and instead subject the AVV amplitude on the right hand side of Eq. (\ref{eq27}) to all possible permutations of indices and momenta, and average over them,  taking into consideration that the AVV amplitude must be computed using the definition
\begin{eqnarray}
\gamma_5=\tfrac{i}{4!}\epsilon^{\mu\nu\alpha\beta}\gamma_\mu\gamma_\nu\gamma_\alpha\gamma_\beta\,.
\label{eq36}\label{eq:def-g_5}\end{eqnarray}
As already mentioned in the 2-d case this is equivalent to a symmetrization of the trace.

 Then, starting with the AAA amplitude, the trace can be rewritten as

\be
T^{AAA}_{\mu\nu\alpha} (p,q,l) &=& \frac{1}{3}\Big( T^{VVA}_{\mu\nu\alpha}(p,q,l) +T^{VVA}_{\mu\alpha\nu}(p,l,q) \nonumber\\
&&+ T^{VVA}_{\alpha\nu\mu}(l,q,p)\Big) \,,\nonumber\\ 
\label{eq31}\label{eq:AAA}\ee
which is consistent with the Bose symmetry of diagram AAA, and does not depend on the value of the AVV anomaly, see Eq. (\ref{eq:AAA-VVA-result})   below.
To obtain the RRR and LLL amplitudes in terms of the amplitudes involving the A and V fields, we start from the algebraic relation between the vertices
 (\ref{eq:vertices}): 
\be
T^{RRR}_{\mu\nu\alpha}(p,q,l)&=&\frac{1}{8}\Big(T^{VVV}_{\mu\nu\alpha}(p,q,l)+T^{VVA}_{\mu\nu\alpha}(p,q,l)+\dots\nonumber\\
&&+T^{AAA}_{\mu\nu\alpha}(p,q,l) \Big)\,,
\label{eq32}\label{eq:AVV+AVA+}\ee

\noindent which contains all the following combinations (keeping fixed their indices and momenta): VVV, VVA, VAV, VAA, AVV, AVA, AAV, AAA. We show that they are equivalent to the expected permutations on indices and momenta of the right hand side of Eq. (\ref{eq27}).
Using the cyclic property of the trace,  Eq. (\ref{eq23}) may be rewritten as
\begin{align} T^{IJK}_{\mu\nu\alpha}(p,q,l) &=&-i\ik \text{Tr}\, \m{V}^{I}_\mu \frac{i}{\slashed{k}+\slashed{k_5}}\m{V}^{K}_\alpha \frac{i}{\slashed{k}+\slashed{k_6}}\m{V}^{J}_\nu \frac{i}{\slashed{k}+\slashed{k_4}} \nonumber\\
&&-i\ik \text{Tr}\,\m{V}^{K}_\alpha \frac{i}{\slashed{k}+\slashed{k_3}} \m{V}^{I}_\mu \frac{i}{\slashed{k}+\slashed{k_1}} \m{V}^{J}_\nu \frac{i}{\slashed{k}+\slashed{k_2}} \,.
\label{eq33}\end{align}

\begin{center}
\begin{figure*}
\includegraphics{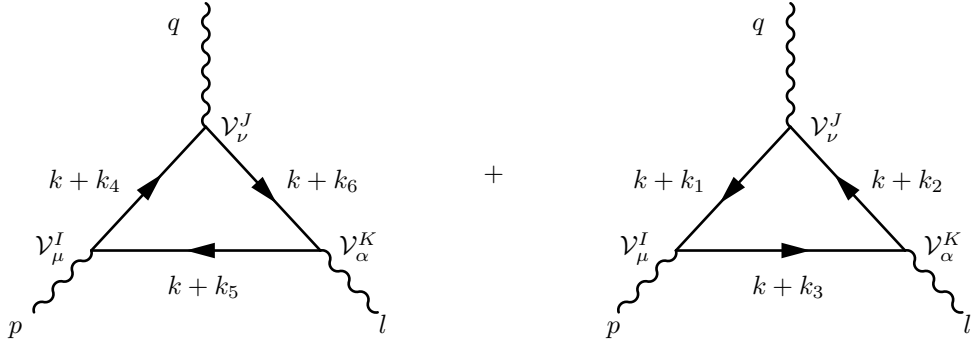}
\caption{The $T^{IKJ}_{\mu\alpha\nu}(p,l,q)$ triangle diagram.}
\label{fig:002}
\end{figure*}
\end{center}

\noindent With the definitions  (\ref{eq24})  this represents the amplitude   $T^{IKJ}_{\mu\alpha\nu}(p,l,q)$ depicted in fig.\ref{fig:002}. Thus one obtains

\be
T^{IJK}_{\mu\nu\alpha}(p,q,l) = T^{IKJ}_{\mu\alpha\nu}(p,l,q)\,. \ee

\noindent In the same way one shows that
\be
T^{IJK}_{\mu\nu\alpha}(p,q,l) = T^{KJI}_{\alpha\nu\mu}(l,q,p)   
\ee
and so
\begin{align}
T^{VAV}_{\mu\nu\alpha}(p,q,l)=T^{VVA}_{\mu\alpha\nu}(p,l,q) \,,\nonumber\\
T^{AVV}_{\mu\nu\alpha}(p,q,l)=T^{VVA}_{\alpha\nu\mu}(l,q,p)\,. 
\label{eq34}
\end{align}
The diagrams involving two axial vertices reduce to the VVV diagram, upon using the definition (\ref{eq:def-g_5}) of $\gamma_5$ in 4-d,
to evaluate the trace containing two $\gamma_5$,
\begin{eqnarray}
\tr [\gamma_\mu\gamma_\nu\gamma_\alpha\gamma_\beta\gamma_5\gamma_\gamma\gamma_\delta\gamma_5]=\tr [\gamma_\mu\gamma_\nu\gamma_\alpha\gamma_\beta \gamma_\gamma\gamma_\delta]
\label{eq37}\end{eqnarray} 
(in this case the result coincides with the trace evaluated using the anti-commutator of $\gamma_5$)
\noindent leading to
\begin{eqnarray}
T^{AAV}_{\mu\nu\alpha}(p,q,l)&=&T^{AVA}_{\mu\nu\alpha}(p,q,l)=T^{VAA}_{\mu\nu\alpha}(p,q,l)\nonumber\\
&=&T^{VVV}_{\mu\nu\alpha}(p,q,l)\,,
\label{eq35}\end{eqnarray}

\noindent with $T^{VVV}_{\mu\nu\alpha}(p,q,l)=0$ by Furry's theorem.
With Eqs. (\ref{eq32}), (\ref{eq:AAA}) and (\ref{eq35}) we obtain finally

\be
T^{RRR}_{\mu\nu\alpha} (p,q,l) &=& \frac{1}{6}\Big( T^{VVA}_{\mu\nu\alpha}(p,q,l) +T^{VVA}_{\mu\alpha\nu}(p,l,q)\nonumber\\
&&+ T^{VVA}_{\alpha\nu\mu}(l,q,p)\Big) \,,
\label{eq38}\label{eq:RRR}
\ee
\noindent and in an analogous way
\be
T^{LLL}_{\mu\nu\alpha} (p,q,l) &=& - \frac{1}{6}\Big( T^{VVA}_{\mu\nu\alpha}(p,q,l) +T^{VVA}_{\mu\alpha\nu}(p,l,q)\nonumber\\
&&+ T^{VVA}_{\alpha\nu\mu}(l,q,p)\Big) \,. 
\label{eq39}\ee

The results (\ref{eq38}) e (\ref{eq39}) are compatible with the Bose symmetry of diagrams RRR and LLL, without the need to fix the value of the anomaly of the amplitude AVV. We emphasize that this achievement relies on the fact that now there is a permutation in the external indices and momenta, as opposed to Eq.  (\ref{eq26}). 

In the following we obtain  the anomalies associated to the left-right currents. Using for instance Eq. (\ref{eq:RRR})  and combining it with the results for the axial and vector anomalies (\ref{eq:AVVanomaly}) yields

\be
\label{NA}
l^\alpha T^{RRR}_{\mu\nu\alpha} (p,q,l) &=&   \frac{1}{6} l^\alpha \Big( T^{VVA}_{\mu\nu\alpha}(p,q,l) +T^{VVA}_{\mu\alpha\nu}(p,l,q) \nonumber\\
&&+ T^{VVA}_{\alpha\nu\mu}(l,q,p)\Big)   \,, 
\ee

\be 
l^\alpha  T^{VVA}_{\mu\nu\alpha}(p,q,l) &=& - \frac{1}{2\pi^2}a\epsilon_{\mu\nu\beta\lambda}p^\beta q^\lambda \,, \nonumber\\
l^\alpha T^{VVA}_{\mu\alpha\nu}(p,l,q) &=&  \frac{1}{4\pi^2}(1+a)\epsilon_{\nu\mu\beta\lambda}p^\beta l^\lambda \nonumber\\
&=& \frac{1}{4\pi^2}(1+a)\epsilon_{\mu\nu\beta\lambda}p^\beta q^\lambda \,,\nonumber\\
l^\alpha T^{VVA}_{\alpha\nu\mu}(l,q,p) &=& -\frac{1}{4\pi^2}(1+a)\epsilon_{\mu\nu\beta\lambda}l^\beta q^\lambda\nonumber\\&=& \frac{1}{4\pi^2}(1+a)\epsilon_{\mu\nu\beta\lambda}p^\beta q^\lambda\,,
\ee
which lead to
\be
l^\alpha T^{RRR}_{\mu\nu\alpha} (p,q,l) &=&  \frac{1}{12\pi^2}\epsilon_{\mu\nu\beta\lambda}p^\beta q^\lambda \,.
\label{eq40}\label{eq:RRRanomaly}\ee 
\noindent 
In a similar way
\be
l^\alpha T^{LLL}_{\mu\nu\alpha} (p,q,l) &=& - \frac{1}{12\pi^2}\epsilon_{\mu\nu\beta\lambda}p^\beta q^\lambda  \,.
\label{eq:LLLanomaly}
\ee

\noindent and
\be
l^\alpha T^{AAA}_{\mu\nu\alpha} (p,q,l) &=& \frac{1}{6\pi^2}\epsilon_{\mu\nu\beta\lambda}p^\beta q^\lambda  
\label{eq41}\label{eq:AAA-VVA-result}\ee
in accordance with the results in \cite{bertlmann-anomalies}.

Let us comment on  Eqs. (\ref{eq:RRRanomaly}),  (\ref{eq:LLLanomaly}), (\ref{eq:AAA-VVA-result}).  The WI of these Bose symmetric amplitudes are not ambiguous.  Like in the 2-d case the average  $\Pi^{VA}_{\mu\nu}+\Pi^{VA}_{\nu\mu}$ was not ambiguous, Eq. (\ref{sd}), the combination over the AVV amplitudes in Eq. (\ref{NA}) is also not, there is an exact cancellation of the arbitrary parameter $a$. But  due to the Furry theorem, the VVV amplitude is absent. This distinguishes the 4-d from the 2-d behavior of the Bose symmetric amplitudes, where an ambiguity appeared in the VV term. Therefore for these WI it is irrelevant whether the trace for the amplitude AVV is calculated using Eq. (\ref{eq:def-g_5}) or the anti-commutator in Eq. (\ref{eq:AAA}). However, and most importantly, the symmetrization of the trace of AVV using Eq. (\ref{eq:def-g_5}) continues to  be mandatory, as it leads to non-vanishing contributions to the anomaly from the amplitudes of mixed chirality as in the 2-d case. These amplitudes are given at the end of this section.
\vspace{0.5cm}

The operator identity (\ref{eq:op3d}) for the vector anomaly is written in terms of the amplitudes $p^\mu T^{VVV}_{\mu\nu\rho}(p,q,l)$, $p^\mu T^{VAV}_{\mu\nu\rho}(p,q,l)$, $p^\mu T^{VAA}_{\mu\nu\rho}(p,q,l)$ and $p^\mu T^{VVA}_{\mu\nu\rho}(p,q,l)$,

\be
p^\mu T^{VVA}_{\mu\nu\rho}(p,q,l) = \frac{1}{4\pi^2}(1+a)\epsilon^{\rho\nu\beta\lambda}l_\beta q_\lambda\,, 
\ee

\begin{eqnarray}
p^\mu T^{VAV}_{\mu\nu\rho}(p,q,l)&=&p^\mu T^{VVA}_{\mu\rho\nu}(p,l,q)\nonumber\\
&=& \frac{1}{4\pi^2}(1+a)\epsilon^{\nu\rho\beta\lambda}q_\beta l_\lambda\nonumber\\
&=&\frac{1}{4\pi^2}(1+a)\epsilon^{\rho\nu\beta\lambda}l_\beta q_\lambda 
\end{eqnarray}
and since the other diagrams do not contribute, one obtains

\be
\langle \partial_\mu J_V^\mu \rangle = \frac{e^2}{16\pi^2}(1+a)\epsilon^{\mu\nu\rho\sigma}F^{A}_{\mu\nu}F^{V}_{\rho\sigma} 
\label{eq42}\ee

To calculate the operator identity for the axial anomaly we need the amplitudes $p^\mu T^{AVV}_{\mu\nu\rho}(p,q,l)$, $p^\mu T^{AAV}_{\mu\nu\rho}(p,q,l)$, $p^\mu T^{AAA}_{\mu\nu\rho}(p,q,l)$  and $p^\mu T^{AVA}_{\mu\nu\rho}(p,q,l)\,.$ From

\be
p^\mu T^{AVV}_{\mu\nu\rho}(p,l,q) = \frac{1}{2\pi^2}a\epsilon^{\nu\rho\lambda\sigma} l_\lambda q_\sigma 
\ee
and
\be
p^\mu T^{AAA}_{\mu\nu\rho}(p,l,q) = \frac{1}{6\pi^2}\epsilon_{\nu\rho\beta\lambda}l^\beta q^\lambda \,,
\ee 
one obtains

\begin{align}
\langle \partial_\mu J_A^\mu \rangle =  \frac{e^2}{16\pi^2}\left(-a \epsilon^{\mu\nu\rho\sigma}F^{ V }_{\mu\nu}F^{ V }_{\rho\sigma} +\frac{1}{3} \epsilon^{\mu\nu\rho\sigma}F^{ A }_{\mu\nu}F^{ A }_{\rho\sigma}\right) \,.
\label{eq43}
\end{align}

In order to calculate  $\langle \partial_\mu J_R^\mu \rangle $ and $\langle \partial_\mu J_L^\mu \rangle $ we need to consider the diagrams $T^{RRR}_{\mu\nu\rho}$, $T^{RLR}_{\mu\nu\rho}$, $T^{RLL}_{\mu\nu\rho}$, etc. An equivalent and easier way is to use the relations among the currents (\ref{eq:currentrelations}) and the fields (\ref{eq20}),

\begin{widetext}
\be
\langle \partial_\mu J_R^\mu\rangle&=&\frac{1}{2}\left(\langle \partial_\mu J_V^\mu \rangle +\langle \partial_\mu J_A^\mu \rangle \right)= \frac{e^2}{32\pi^2}\Bigg(\dfrac{1}{3} \epsilon^{\mu\nu\rho\sigma}F^{R}_{\mu\nu}F^{R}_{\rho\sigma}-\frac{1}{2}\left(a+\dfrac{1}{3}\right)\epsilon^{\mu\nu\rho\sigma}\left(F^{L}_{\mu\nu}F^{L}_{\rho\sigma}+F^{R}_{\mu\nu}F^{L}_{\rho\sigma}\right) \Bigg)\,,\nonumber\\
\langle \partial_\mu J_L^\mu\rangle&=&\frac{1}{2}\left(\langle \partial_\mu J_V^\mu \rangle -\langle \partial_\mu J_A^\mu \rangle \right) = -\frac{e^2}{32\pi^2}\Bigg(\dfrac{1}{3} \epsilon^{\mu\nu\rho\sigma}F^{L}_{\mu\nu}F^{L}_{\rho\sigma}-\frac{1}{2}\left(a+\dfrac{1}{3}\right)\epsilon^{\mu\nu\rho\sigma}\left(F^{R}_{\mu\nu}F^{R}_{\rho\sigma}+F^{R}_{\mu\nu}F^{L}_{\rho\sigma}\right) \Bigg)\,.
\label{eq44}\label{eq:0010} \ee
\end{widetext}

\noindent Nevertheless it is instructive to check these results going through the evaluation of diagrams  RRR, RLL, RLR, RRL, LLL, LLR, LRL and LRR; it is important to realize that if one uses the identity (\ref{eq7}) only RRR and LLL do not vanish, in contradistinction to the approach advocated in the present work, which uses instead the symmetrization of the trace and leads to a non vanishing result for all these amplitudes.  With the same prescription as in  eq. (\ref{eq:AVV+AVA+}), we obtain

\begin{widetext}
\begin{eqnarray}
T^{RLL}_{\mu\nu\alpha}(p,q,l)&=&\frac{1}{8}\left(-T^{VVA}_{\mu\nu\alpha}(p,q,l)-T^{VAV}_{\mu\nu\alpha}(p,q,l)+T^{AVV}_{\mu\nu\alpha}(p,q,l)+T^{AAA}_{\mu\nu\alpha}(p,q,l) \right)\nonumber\\
&=& \frac{1}{8}\bigg(-T^{VVA}_{\mu\nu\alpha}(p,q,l)-T^{VVA}_{\mu\alpha\nu}(p,l,q)+T^{VVA}_{\alpha\nu\mu}(l,q,p)+\frac{1}{3}\Big(T^{VVA}_{\mu\nu\alpha}(p,q,l)+T^{VVA}_{\mu\alpha\nu}(p,l,q)\nonumber\\
&&+T^{VVA}_{\alpha\nu\mu}(p,q,l) \Big) \bigg) = \frac{1}{6}T^{VVA}_{\alpha\nu\mu}(l,q,p)-\frac{1}{12}\left(T^{VVA}_{\mu\nu\alpha}(p,q,l)+T^{VVA}_{\mu\alpha\nu}(p,l,q)\right) 
\end{eqnarray}
\end{widetext}

\noindent which leads to

\begin{eqnarray}
p^\mu T^{RLL}_{\mu\nu\alpha}(p,q,l)&=& \frac{1}{8\pi^2}(a+\frac{1}{3})\epsilon_{\nu\alpha\sigma\lambda}l^\sigma q^\lambda \,;
\label{eq45}\end{eqnarray}

and a\noindent nalogously,
\begin{eqnarray}
T^{RLR}_{\mu\nu\alpha}(p,q,l)&=&\frac{1}{8}\Big(T^{VVA}_{\mu\nu\alpha}(p,q,l)-T^{VAV}_{\mu\nu\alpha}(p,q,l)\nonumber\\
&&+T^{AVV}_{\mu\nu\alpha}(p,q,l)-T^{AAA}_{\mu\nu\alpha}(p,q,l) \Big)\nonumber\\
&=& \frac{1}{12}\left(T^{VVA}_{\mu\nu\alpha}(p,q,l)+T^{VVA}_{\alpha\nu\mu}(l,q,p)\right)\nonumber\\
&&-\frac{1}{6}T^{VVA}_{\mu\alpha\nu}(p,l,q)  
\end{eqnarray}

\noindent implies
\begin{eqnarray}
p^\mu T^{RLR}_{\mu\nu\alpha}(p,q,l) = \frac{1}{16\pi^2}\left(a+\frac{1}{3} \right)\epsilon_{\nu\alpha\sigma\lambda}l^\sigma q^\lambda \,,
\label{eq46}\end{eqnarray}
and
\begin{eqnarray}
T^{RRL}_{\mu\nu\alpha}(p,q,l)&=&\frac{1}{8}\Big(-T^{VVA}_{\mu\nu\alpha}(p,q,l)+T^{VAV}_{\mu\nu\alpha}(p,q,l)\nonumber\\
&&+T^{AVV}_{\mu\nu\alpha}(p,q,l)-T^{AAA}_{\mu\nu\alpha}(p,q,l) \Big)\nonumber\\
&=& \frac{1}{12}\Big(T^{VVA}_{\mu\nu\alpha}(p,q,l)+T^{VVA}_{\alpha\nu\mu}(l,q,p)\Big)\nonumber\\
&&-\frac{1}{6}T^{VVA}_{\mu\alpha\nu}(p,l,q)  
\end{eqnarray}
leads to
\begin{eqnarray}
p^\mu T^{RRL}_{\mu\nu\alpha}(p,q,l) = \frac{1}{16\pi^2}\Big(a+\frac{1}{3}\Big)\epsilon_{\nu\alpha\sigma\lambda}l^\sigma q^\lambda \,.
\label{eq47}\end{eqnarray}

\noindent 
Finally, with the results of Eqs. (\ref{eq45}), (\ref{eq46}) and (\ref{eq47}) we obtain the first equation displayed in (\ref{eq:0010}); and the same can be done for the second equation.

A comment is in order regarding the four-divergence of the currents (\ref{eq:0010}) and its general character in the description of the anomalies.  

Were the currents to be obtained from the sum of two Weyl spinor anomalies, each obtained from an unambiguous anomalous contribution of a single left-  or a single right- Weyl spinor (denoted in the literature as the consistent anomaly or left-right symmetric anomaly)\cite{Hill2006,Bardeen}, it would be reproduced from  (\ref{eq:0010}) setting $a=-\frac{1}{3}$. Note the subtle difference to the case of a single left (right) Weyl spinor anomaly (not their sum); then the  four-divergence of the currents in  (\ref{eq:0010}) should not depend on the arbitrary value $a$, as this would imply that the single currents would be ambiguous. Indeed the single currents are unambiguous,  since, for example,  the right fields are absent from the beginning from the evaluation of the left current,  meaning that one should set them to zero in the expression for the four-divergence of the left current in (\ref{eq:0010}), and the opposite for the right current. This elliminates any dependence on $a$ and delivers the correct value of the anomaly.

Rephrased in terms of the triangle contributions, the left-right consistent anomaly corresponds to consider only the sum of the RRR and LLL pieces in the evaluation of the currents. By now we have understood that these, in turn,  are the only ones that result from the calculation of the traces using the anticommutator of the $\gamma_5$ matrix. We have demonstrated that not abiding by this property, and instead symmetrizing the trace, the general result displayed in (\ref{eq:0010}) is tantamount to have in addition the non-vanishing contributions of the RRL, RLR, LLR and LRL amplitudes. 

The left-right symmetric anomaly is lifted if one considers the coupling of the left and right Weyl spinors taken together with the left  and right fields \cite{Hill2006}, resulting in an ambiguity in the form of the ABJ anomaly. The vector current conservation is   implemented in \cite{Hill2006}  by the addition of appropriate counterterms. In our approach the result (\ref{eq:0010}) allows to fix the anomaly either in the vector ($a=-1$), or in the axial ($a=0)$  currents, or to distribute it symmetrically ($a=-\frac{1}{3}$).   \\  

\section{Concluding remarks}\label{sec4}

The purpose of the present contribution was to show that the Bose symmetry of amplitudes in the left-right symmetry representation of gauge fields or currents does not  impose any new restriction  on the values that the corresponding left and right chiral anomalies may take, and consequently on the values of the related vector and axial-vector anomalies. Examples in the literature that alert to these Bose symmetry related restrictions can be found for instance  in the textbooks \cite{Pokorski}  and  \cite{bertlmann-anomalies}. These extra constraints lead to an equal distribution of the anomaly in the vector and axial WI. They are thus inconsistent with the expected values that the vector and axial anomalies must fulfill, if for instance  invariance of the gauge current (or alternatively of the axial current) is to be implemented.

Using IReg we have argued that such restrictions are an artifact of inadvertently using the anticommuting property of the $\gamma_5$ matrix (and its analogue in two dimensions) within divergent integrals. We have analyzed two case studies, resorting to the chiral Schwinger model in 2-d and the Adler-Bell-Jackiw
triangle anomaly in 4-d, in the equivalent left - right and vector - axialvector representations. We have repeatedly shown that the use of the anti-commutator in IReg is problematic even when staying in the physical dimension of the problems addressed. In practical terms the use of the anti-commutator in divergent integrals is translated in IReg either into fixing the values of some ({\it a priori} arbitrary) surface terms to zero and consequently dismissing potential contributions to the value of an anomaly, or in generating spurious finite but fixed contributions.  

We have demonstrated that with a minimal prescription to deal with the $\gamma_5$ matrix, which resides in using its basic definition (\ref{eq36}), 
together with the IReg technique, it is possible to obtain a result that respects Bose symmetry independently of the values  of the anomalies.  
For example one of the results obtained by using our minimal prescription concerning the  left- and right-handed currents in the 4-d case, is that the RLR, RLL, LRL and LRR diagrams have non-vanishing contributions to the anomaly, in the form of finite (arbitrary) surface terms (embodied in the parameter $a$), eq. (\ref{eq44}),  as opposed to the case where the anti-commutator has been used, where only the RRR and LLL diagrams survive. These arbitrary parameters leave open the distribution of the values of the anomalies until the very end of the calculation, to be then fixed according to the Physics of the problem considered.  This has the obvious advantage that all symmetries can be discussed on equal footing within the same framework. In particular it lead us to the appealing conclusion that Bose symmetry does not interfere with the anomalous breaking of dynamical symmetries, in the examples considered.  The desired equivalence of representing the anomaly in terms of the  left and right currents (\ref{eq44}) or in terms of the vectorial (\ref{eq42}) and axial currents (\ref{eq43}) is manifestly implemented.   
 
We would like to emphasize that our conclusion of Bose symmetry not giving extra constraints on how to distribute the anomaly, if left- and right currents are used instead of vector and axial currents, is a result that should be reproduced by other methods as well. We used IReg to demonstrate it, since it does not resort to dimensional schemes and due to its convenient “democratic” display of the anomalies, while preserving fundamental principles such as unitarity, Lorentz invariance and causality. We should however keep in mind that in other methods the conditions may differ from the ones used in IReg. In particular IReg uses always the cyclicity of the trace.  On the other hand there exist approaches that show a correspondence between noncyclicity of the trace and anomalies \cite{Kreimer} and rely on the anticommuting property of $\gamma_5$, see also \cite{Jegerlehner}.  It would be interesting to understand whether such approaches are also free from any extra conditions stemming from Bose symmetrization of the left-right currents impacting on the values that the anomaly takes. 

\section{Acknowledgements}

J.S.P. and M.S. acknowledge financial support by  CNPq (Conselho Nacional de Desenvolvimento Científico e Tecnológico) - Brazil. J.S.P. and A.L.C. acknowledge financial support by CAPES (Coordenação de Aperfeiçoamento de Pessoal de Nível Superior) - Brazil. B. Hiller acknowledges partial support from the FCT (Portugal) project UID/FIS/04564/2016.

\end{document}